\documentclass{aa}
\usepackage{psfig}

\newcommand{\cmcub}{~cm$^{-3}$}

\newcommand{\msun}{\ifmmode M_{\odot} \else M$_{\odot}$\fi}
\newcommand{\rsun}{\ifmmode R_{\odot} \else R$_{\odot}$\fi}
\newcommand{\lsun}{\ifmmode L_{\odot} \else L$_{\odot}$\fi}
\newcommand{\zsun}{\ifmmode Z_{\odot} \else Z$_{\odot}$\fi}

\newcommand{\WHa}{\ifmmode {EW(\rm H}\alpha) \else EW(H$\alpha$\fi)}
\newcommand{\WHb}{\ifmmode {EW(\rm H}\beta) \else EW(H$\beta$\fi)}
\newcommand{\Ha}{H$\alpha$}
\newcommand{\Hb}{\ifmmode {\rm H}\beta \else H$\beta$\fi}

\newcommand{\hii}{H~{\sc ii}}

\newcommand{\Nii}{[N~{\sc ii}] $\lambda$6584}

\newcommand{\Oii}{[O~{\sc ii}] $\lambda$3727}
\newcommand{\oii}{[O~{\sc ii}]}
\newcommand{\Oiii}{[O~{\sc iii}] $\lambda$5007}

\newcommand{\oiii}{[O~{\sc iii}]}

\newcommand{\sii}{[S~{\sc ii}]}

\newcommand{\Ostrl}{$O_{23}$}

\def\apj{ApJ}

\def\apjs{ApJS}

\begin{document}

\thesaurus{08(11.01.1;11.05.2;11.09.4;11.19.5;09.08.01;
    )}

\title{
Global emission line trends in spiral galaxies: the reddening and
metallicity sequences}
\author{ Gra\.{z}yna Stasi\'nska\inst{1}
\and Laerte Sodr\'e Jr. \inst{2}}

\offprints{grazyna.stasinska@obspm.fr}
\institute{DAEC, Observatoire
de Paris-Meudon, 92195 Meudon Cedex, France
(grazyna.stasinska@obspm.fr)
\and Departamento de Astronomia, Instituto Astron\^omico
                e Geof\'\i sico da USP, Av. Miguel Stefano 4200, 04301-904
                S\~ao Paulo, Brazil \\
                (laerte@iagusp.usp.br)}

\date{Received / Accepted }

\titlerunning{The reddening and metallicity sequence in spiral galaxies}
\authorrunning{G. Stasi{\'n}ska \& L. Sodr\'e Jr.}

\maketitle


\begin{abstract}

We have explored the  emission line trends in the integrated spectra
of normal spiral galaxies of the Nearby Field Galaxy Survey
(Jansen et al. 2000a and b), in order to investigate the relationships between
dust extinction, metallicity and some macroscopic properties of spiral
   galaxies.

We found a very strong correlation between the \Hb\ and \Ha\ equivalent widths,
implying that the difference between
the extinction of the stellar and the nebular light
   depends only on the intrinsic colours of the galaxies,
being larger for redder galaxies.

The usual  metallicity indicator for giant \hii\ regions
   ([O~{\sc iii}] $\lambda$4959,5007 + [O~{\sc ii}] $\lambda$3726,3729)
/ H$\beta$
is not appropriate for integrated spectra of spiral galaxies,
probably due to metallicity gradients.
Much better qualitative metallicity indicators are found to be
[N~{\sc ii}] $\lambda$6584/[O~{\sc ii}] $\lambda$3726,3729 and
[N~{\sc ii}] $\lambda$6584/\Ha,  the
latter having the advantage of being independent of reddening and being
applicable also for galaxies with weak emission lines.

With these indicators, we find that the nebular extinction as derived
from the Balmer decrement
strongly correlates  with the effective metallicity
of the emission line regions.

The overall metallicity of the emission line regions
   is  much better correlated  with galaxy colours
than with morphological types.

A Principal Component Analysis on a 7-D parameter space showed that the
   variance
is produced, in first place, by the metallicity and parameters linked to the
stellar populations,
   and, in second place,
by the surface brightness, which is linked to the dynamical
history of the galaxies. The absolute magnitude, related to the
mass of the galaxy,  comes only in the third place.

\end{abstract}

\section{Introduction}

The interpretation of integrated spectra of nearby galaxies becomes
increasingly important with the profusion of studies of galaxies in
clusters (e.g., Caldwell et al., 1993; Dressler et al. 1999; Poggianti et al.
1999; Balogh et al. 1999)
and galaxies at large redshifts
(e.g., Colless et al. 1990; Hammer et al. 1997; Lilly et al. 1998,
Guzman et al. 1998).

Indeed, for the studies of distant galaxies, the spatial
resolution obtained with ground based instruments makes any
morphological classification difficult, and even from space, morphological
studies are bound to be crude.  Therefore, most information on distant
galaxies will come from photometry or spectroscopy, and the sampled
regions will cover a significant part - if not all - of the galaxy
surface.

Techniques have been devised to classify  galaxies using their
spectra only (Morgan \& Mayall; 1957, Sodr\'e \& Cuevas; 1994, 1997;
Connolly et al. 1995; Zaritsky, Kennicutt \& Huchra 1994; Folkes et 
al. 1996; Galaz
\& de Lapparent 1998; Sodr\'e \& Stasi\'nska 1999). It has been shown
that the  results
correlate  rather  well  with  the  galaxy  morphological  types.  The
spectral classification actually presents some advantages even for the
study of nearby galaxies: whatever  criteria are adopted, they allow an
objective and  continuous classification of the  galaxies, in contrast
with the conventional morphological classification which is by nature
subjective
and provides discrete classes. In addition, any interpretation in terms
of stellar content and evolution is directly related with the galaxy
spectra, while the link with galaxy morphology is rather elusive.

In a former paper (Sodr\'e \& Stasi\'nska 1999, hereinafter SS99), based on the
integrated spectra for normal spiral galaxies from the Kennicutt
(1992) atlas, we have shown that emission line ratios and equivalent
widths  are impressively well
correlated with the galaxy spectral type obtained from a Principal
Component Analysis of the continuum and the absorption features. The
correlation is far better with galaxy spectral types than with
morphological types. We also found that the extinction derived from
the emission line Balmer decrement
correlates with the galaxy spectral type, being larger for early type
galaxies.

But the sample was small: there were only 15 {\em normal} spirals in the
Kennicutt (1992) atlas. In the meantime, Jansen et al. (2000a and 2000b)
published a comprehensive photometric and spectrophotometric atlas of
200 nearby field galaxies (the Nearby Field Galaxy Survey, hereinafter NFGS).
   The objects were selected so as to span the full
range in galaxy luminosities and morphological types. As in Kennicutt
(1992), the integrated galaxy spectra were obtained by drifting a long
slit across the galaxy. This considerably richer and more
appropriate data base motivated us to revisit the study of SS99,
in order to see whether the formerly found trends
are confirmed, and to refine their empirical interpretation.

In this paper, we essentially investigate the overall effective
extinction and metallicity of spiral galaxies through an empirical
analysis of the properties of their global emission lines. We also
perform a Principal Component Analysis in order to understand the
relation amongst the variables describing the galaxies.

As in SS99, our paper deals only with normal galaxies. Therefore, we
have rejected from the NFGS sample all the
galaxies with Seyfert nuclei and nuclear starbursts, as well as the
galaxies with indication of close interaction or strong peculiarity,
as indicated in the appendix of Jansen et al. (2000a). In the
following, the expression NFGS sample will refer to the 158 galaxies
that have no evident sign of peculiarity.

All the galaxies in our sample are found at redshifts lower
than 0.05. Their blue absolute magnitudes
range between -22 and -13. For most of the diagrams
investigated in this paper, relevant data are available for 100--120
objects.

\section{Dust extinction}

The dust content of galaxies is a controversial subject. While the first
comprehensive analysis of this problem (Holmberg 1958, 1975; de Vaucouleurs
1959) have proposed that spiral disks are largely
transparent, more recent works have suggested that these disks are opaque
or at least far more obscured than previously thought (cf. Valentijn 1990;
Disney, Davies \& Phillipps, 1989; James \& Puxley 1993).
So far, there is still no consensus  on whether
galaxies are optically thick or thin (see Wang \& Heckman 1996 or
Meurer \& Seibert 2001
and references therein).
Actually, the galaxy extinction
depends not only on the amount of dust and its composition, but also on
the distribution of dust relative to the light sources. This point too
has been disputed. For instance, some authors suggest a foreground
screen dust geometry (Calzetti, Kinney \& Storchi-Bergmann 1994, 1996; but see
Witt, Thronson \& Capuano 1992), while
others propose hybrid models, with the
dust partially distributed in a foreground screen and partially
concentrated in the star forming regions (e.g. Charlot \& Fall 2000).
Indeed, Charlot \& Fall (2000; see also Charlot \& Longhetti 2001)
present a model for the effects of dust on
integrated galaxy spectra that explains several observational
properties of starburst galaxies, showing that there is a sequence in
the overal dust content of the galaxies. Radial gradients should also
play a role, and Giovanelli et al. (1994) and Peletier et al. (1995)
propose that the dust distribution presents a strong radial dependence,
with considerable obscuration near the centers of galaxies and little
at the outer edges (see also Nelson, Zaritsky and Cutri 1998).

In nearby starburst galaxies, there is evidence from bidimensional
spectroscopy that stars, gas and dust are decoupled (Ma\'\i
z-Apell\'aniz et al. 1998)
and  the attenuation inferred from the
H$\alpha$/H$\beta$ ratio is typically higher than that inferred from
the spectral continuum (e.g., Fanelli, O'Connell \& Thuan 1988;
Calzetti, Kinney \& Storchi-Bergmann 1994; Mayya \& Prahbu 1996;
Calzetti 1997).
Recently, Poggianti et al. (1999) and Poggianti \& Wu (2000) have
shown that the so-called e(a) spectrum can be reproduced assuming
that in a current starburst the dust extinction affects the young
stellar populations much more than the older stars. This effect has been
called selective dust extinction by Poggianti and collaborators,
and is a consequence of the fact that a large fraction -- but not
all --
of the dust in galaxies is associated with star formation regions, absorbing
a significant fraction of the light emitted by the young stars.
We now present an empirical evidence
that selective extinction
is indeed affecting the integrated spectra of spiral galaxies.

\subsection{Selective Extinction}

\begin{figure}
\centerline{\psfig{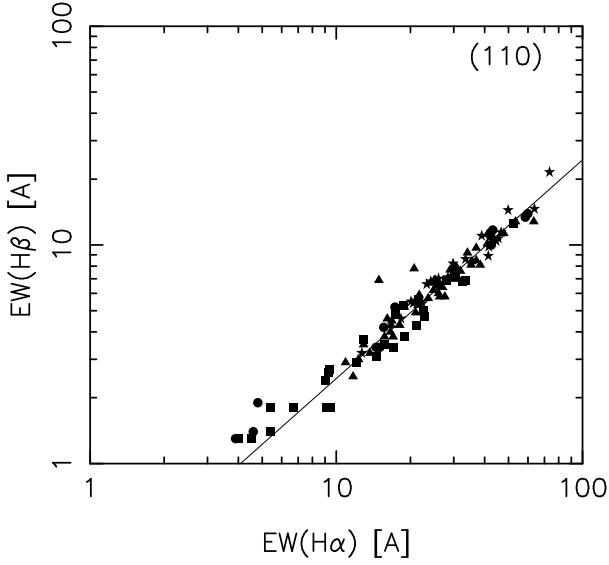}}
\caption{ \WHb\ as a function of \WHa\ for the normal spiral galaxies
from the NFGS sample. Different symbols correspond to galaxies of
different morphological types $T$.
Circles: $-5 \leq T \leq 0$; squares: $1 \leq T \leq 3$;
triangles: $4 \leq T \leq 6$; Circles: $7 \leq T \leq 10$.
The total number
of the objects appearing in the diagram is given in parenthesis.
Such a presentation is adopted for all the observational diagrams in
this paper.
The \WHb\ = 0.245 \WHa\ line is shown.
}
\end{figure}

\begin{figure}
\centerline{\psfig{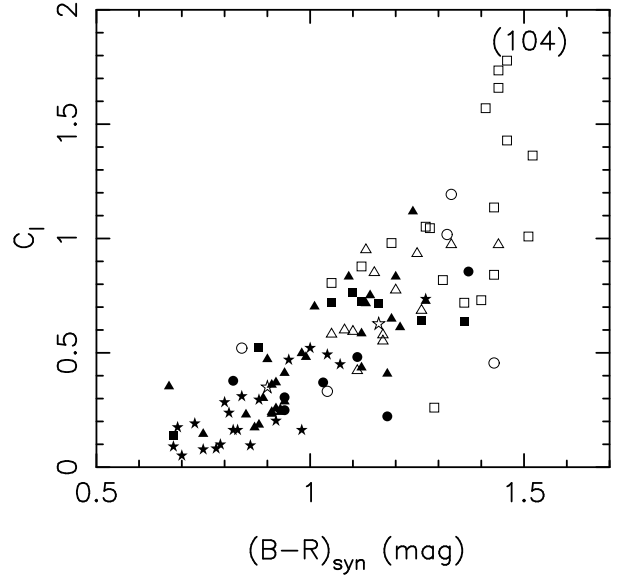}}
\caption{$C_l$ as a function of the galaxy synthetic colour $(B-R)$
derived from the spectra.
Symbols have the same meaning as in Fig. 1. Open symbols correspond to galaxies
with \WHa $<$ 10~\AA\ or  \WHb $<$ 5~\AA.
}
\end{figure}

Fig. 1 shows the emission line equivalent width of \Hb, \WHb, as a
function of that of \Ha, \WHa, for the  110
galaxies from the NFGS
sample with available data. The shape of the symbols correspond to
different morphological classes, as indicated in the figure
caption. As explained by Jansen et al. (2000b), these equivalent
widths were extracted from the spectra by placing the limits of the
measurement window well inside the absorption trough, thus minimizing
the effect of the underlying stellar absorption, which becomes
appreciable at small equivalent widths. Following the
prescription of Jansen, Franx \& Fabricant (2001),
we have applied
an additional correction for residual absorption (1.5~\AA\ for \Ha\
and  1~\AA\ for \Hb).

The correlation between \WHb\ and \WHa\ is impressively strong and is well
represented by a relation of the type
\begin{equation}
EW({\rm H}\beta) = A \times EW({\rm H}\alpha)
\end{equation}
where $A$ is a constant. Using the
ordinary least squares bisector estimation for the fitting
(Isobe et al. 1990), we find $A = 0.245 \pm 0.007$.
This correlation is even better than the one found by SS99 for the
Kennicutt sample.
The uncertainties in the emission line equivalent widths may be
appreciable for \WHa\ or \WHb\ smaller than, say, 10~\AA\, but the
value of the slope $A$
is essentially determined by larger equivalent widths and is
not sensitive to these uncertainties.
We have performed a non-parametric  Kendall's $\tau$
statistical test
(Press et al. 1992) to evaluate quantitatively the quality of the
fitting, for the whole sample and for the sub-sample with \WHa\ $> 5$\AA~
and \WHb\ $> 10$\AA. The results are presented in Table 1 and reveal
that the correlation between  \WHa\ and \WHb\ is indeed highly significant for
both samples.
For an easy consultation, the results of this test for all the
observational diagrams investigated in this study
have been summarized in Table 1. In this table we present, for each plot
and sample, the number of data points, the value of the non parametric
correlation
coefficient $\tau$, and its two-sided significance level, $prob$. Note that
small values for $prob$ indicates that the correlation or
anti-correlation between the variables is significant.

We found that the residuals in the \WHb\ vs \WHa\ relation are
independent of the galaxy ellipticity (therefore inclination) as well
as of the galaxy absolute magnitude (and any combination of
ellipticity and magnitude). They are also independent of galaxy
morphological type.

The correlation shown in Fig. 1 has important implications concerning
the nature of the internal
reddening of the galaxies, in particular in terms of a link between the
effective extinction in the continuum and in the emission lines. Let us write
an emission line equivalent width as a ratio between the
emission line intensity $I$ and the flux $F_c$ at the adjacent continuum
in the observed spectrum: $EW = I/F_c$.
The relation between the
observed and intrinsic (i.e., extinction corrected) intensities of
H$\alpha$ and H$\beta$ is
\begin{equation}
\frac{I({\rm H}\alpha)}{I({\rm H}\beta)}= \delta  \times
10^{-C_l(f({\rm H}\alpha)-f({\rm H}\beta))}
\end{equation}
where $\delta$ is the intrinsic intensity ratio of these two lines,
$C_l$ is a measure of the effective extinction at H$\beta$ in the region
producing the lines, and $f(\lambda)$ is a function that gives the
wavelength dependence of the extinction.
Let $C_c$ be a measure of the
effective amount of extinction at the wavelength of H$\beta$
in the region producing the continuum, so that the observed and intrinsic
fluxes (denoted by $F_c$ and $F_c^0$, respectively)
at a given wavelength are related by
\begin{equation}
   F_c(\lambda) = F_c^0(\lambda) \times 10^{-C_c f(\lambda)}
\end{equation}
Eq. 1 implies that the effective extinction at H$\beta$ is
related to the ratio of the
observed continuum fluxes at H$\alpha$ and H$\beta$ by
\begin{equation}
   C_l = \frac{-1}{f({\rm H}\alpha)-f({\rm H}\beta)}\log \left( 
\frac{1}{A\delta}
\frac{F_c({\rm H}\alpha)}{F_c({\rm H}\beta)} \right)
\end{equation}
and that the difference between the line and continuum effective
extinctions is related to the ratio of the intrinsic continuum fluxes by
\begin{equation}
C_l - C_c = \frac{-1}{f({\rm H}\alpha)-f({\rm H}\beta)}\log \left(
\frac{1}{A\delta}
\frac{F_c^0({\rm H}\alpha)}{F_c^0({\rm H}\beta)} \right)
\end{equation}
Since $F_c^0({\rm H}\alpha)/F_c^0({\rm H}\beta)$ depends only on the
stellar populations
of the galaxy, Eq. 5 is indeed  an empirical prescription
for the relation between selective dust extinction
and the galaxy stellar populations.

Although we have no way to determine the value of $C_c$
from the available data, some implications of Eq. 5 may be inferred.
First, the
difference between $C_l$ and $C_c$ is essentially dependent on the intrinsic
colour of the galaxy and is larger for intrinsically redder
galaxies. Second, if $C_l \ge C_c$, a lower limit for the
intrinsic continuum flux ratio at the wavelengths of H$\alpha$ and
H$\beta$ is $F_c^0({\rm H}\alpha)/F_c^0({\rm H}\beta)=A \delta$.
Assuming $\delta= 2.9$ (the case B recombination value corresponding
to a temperature of 9000K), and using the Seaton (1979) extinction law, we
find that $F_c^0({\rm H}\alpha)/F_c^0({\rm H}\beta)$ in spiral galaxies
should be larger than 0.70.

We have used the software P\'EGASE.2 (Fioc \&
Rocca-Volmerange 1997) with a Salpeter IMF and default values to
compute this ratio for a few simple model galaxy spectra.
For an instantaneous burst
15 Gyr old we obtain $F_c^0({\rm H}\alpha)/F_c^0({\rm H}\beta)=0.84$
and Eq. 5 implies $C_l-C_c=0.21$, whereas for a model with
constant star formation rate and same age this ratio is 0.75 and Eq. 5
leads to $C_l-C_c=0.07$. Additional clues on $C_c$ may be obtained
from spectral synthesis methods that include explicitly in the analysis
the effects of extinction on the spectra
(e.g., Cid Fernandes et al. 2001).

\begin{table*}
\begin{center}
\caption{Results of Kendall's $\tau$ non parametric test for the
observational diagrams.}
\begin{tabular}{c|ccc|ccc}
\hline
\multicolumn{1}{c}{figure number}&\multicolumn{3}{c|}{whole sample}&
\multicolumn{3}{c}{
sample with EW(H$\beta$) $>$5 \AA~ and EW(H$\alpha$) $>$10 \AA} \\
  & N & $\tau$ &  $prob$ & N & $\tau$ &  $prob$ \\
\hline
    1 & 110 &   8.79E-01 &  3.69E-42  &   70 &  7.93E-01 & 2.66E-22 \\
    2 & 104 &   6.69E-01 &  8.05E-24  &   66 & 6.28E-01 & 8.77E-14 \\
    3 & 104 &  -2.24E-01 &  7.71E-04  &   66 & -4.20E-01 & 6.00E-07 \\
    6 & 104 &   6.00E-01 &  1.69E-19  &   66 & 6.01E-01 & 9.90E-13 \\
    7 & 104 &   4.09E-01 &  7.37E-10  &   66 & 5.34E-01 & 2.23E-10 \\
   8a & 116 &  -5.50E-01 &  2.06E-18  &   70 & -2.53E-01 & 1.96E-03 \\
   8b & 123 &  -5.27E-01 &  5.74E-18  &   70 & -2.40E-01 & 3.31E-03 \\
   8c &  81 &  -5.24E-01 &  4.40E-12  &   64 & -4.99E-01 & 5.65E-09 \\
   8d & 116 &  -2.34E-02 &  7.09E-01  &   68 &  4.24E-01 & 3.19E-07 \\
   9a & 116 &   3.40E-01 &  6.28E-08  &   70 &  5.31E-03 & 9.48E-01 \\
   9b & 123 &   3.44E-01 &  1.72E-08  &   70 &  1.09E-01 & 1.81E-01 \\
   9c &  81 &   2.29E-01 &  2.44E-03  &   64 &  1.68E-01 & 5.02E-02 \\
   9d & 116 &  -2.23E-03 &  9.72E-01  &   68 & -3.75E-01 & 5.96E-06 \\
10a  & 94  & -4.78E-01 &  8.70E-12   &   70 & -5.20E-01 & 1.87E-10 \\
10b  & 94  &  2.63E-01 &  1.75E-04   &   70 & 2.44E-01  & 2.85E-03 \\
10c  & 108 &   4.37E-01 &  2.12E-11  &   70 & 5.83E-01  & 9.54E-13 \\
10d  & 106 &   6.36E-01 &  4.19E-22  &     68 &  6.32E-01 &  2.61E-14 \\
11a  & 108 &  -5.38E-01 &  1.49E-16  &     70 &  -5.87E-01  &   6.47E-13 \\
11b  & 108 &  -5.03E-01  &  1.15E-14  &     70 &  -5.38E-01  &  4.34E-11 \\
11c  & 108 &   3.36E-01 &  2.53E-07  &     70  &  3.84E-01  &  2.66E-06 \\
\hline
\end{tabular}
\end{center}
\end{table*}

\subsection{The extinction of the nebular light}

We obtained $C_l$ from the observed \Ha/\Hb\ ratio for the 110 normal
galaxies of the NFGS sample with available data.

Fig. 2 plots
$C_l$ as a function of the galaxy color $(B-R)$.  The open symbols
correspond to galaxies
with \WHa $<$ 10~\AA\ or  \WHb $<$ 5~\AA\, for which the value of
$C_l$ tends to be more uncertain.
A very good correlation is seen,
with $C_l$ larger  for redder objects  (the correlation coefficient $\tau$
shown in Table 1 is larger than 0.6).
This behavior is indeed expected from Eq. 4.  As
already shown by Jansen, Franx \& Fabricant (2000), there is a
correlation between $C_l$ and the galaxy blue absolute magnitude,
although the dispersion is much larger than for the color.  $C_l$ also
correlates with the morphological types, decreasing towards later
types, confirming the trend found in SS99.  The maximum extinction is
observed for galaxies Sb, but the variance is large. On the other
hand, there is no correlation whatsoever with the galaxy apparent
ellipticity, implying that at least part of the obscuring dust is
concentrated close to the emitting sources.

We now examine whether there is a relation between $C_l$
and the overall metallicity of the galaxies.
So far, there have been contradictory claims in this respect.
Zaritsky, Kennicutt \& Huchra (1994) found no evidence for a 
systematic dependence
   between reddening and abundance in a sample of 39 disk galaxies. In
their study,
the characteristic metallicity of the galaxies was derived from the
value of the
metallicity indicator
   ([O~{\sc iii}] $\lambda$5007,4959 + [O~{\sc ii}] $\lambda$3726,3729)
/ H$\beta$
(hereinafter \Ostrl)
(Pagel et al. 1979, McGaugh 1991)  at the isophotal radius,
interpolated from the values observed in giant \hii\ regions.
In other contexts, however, it has been suggested
that the extinction derived from the Balmer decrement might be
related to metallicity.
For example,
Campbell, Terlevich \& Melnick (1986) found a loose correlation
between the extinction and the oxygen abundance in a sample of \hii\ galaxies.
Also,  invidivual giant \hii\ regions in
spiral galaxies exhibit a decrease of extinction with galactocentric radius
   (Kennicutt \& Garnett 1996; Van Zee et al. 1998a), but this is not a
general rule
(Martin \& Roy, 1992 and references therein).

\begin{figure}
\centerline{\psfig{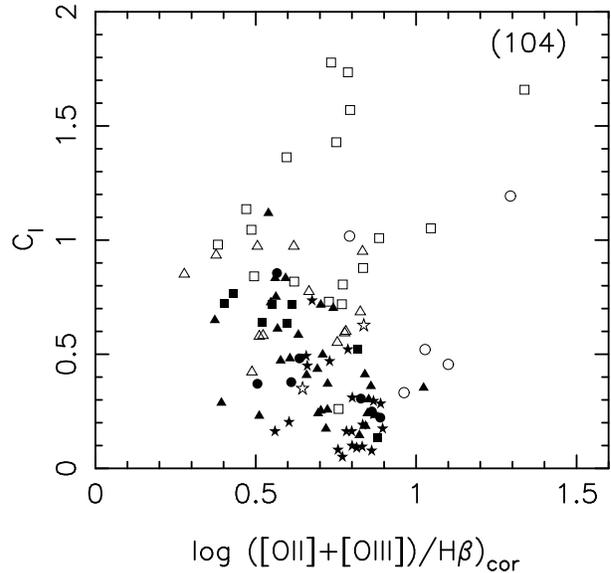}}
\caption{$C_l$ as a function of
[O~{\sc iii}] $\lambda$5007,4959 + [O~{\sc ii}] $\lambda$3726,3729) / H$\beta$
corrected for reddening. Same symbols as in Fig. 2.
}
\end{figure}

\begin{figure}
\centerline{\psfig{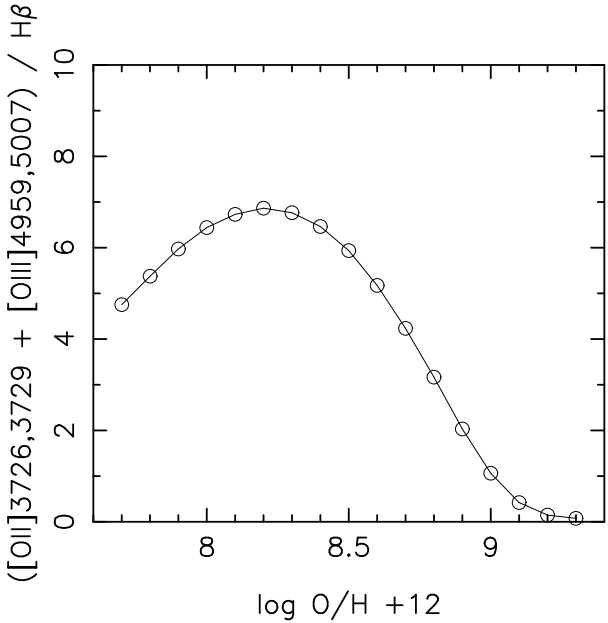}}
\caption{ [O~{\sc iii}] $\lambda$5007,4959 + [O~{\sc ii}]
$\lambda$3726,3729) / H$\beta$
   as a function of O/H for a series of  photoionization models
representative of  giant \hii\ regions
(total number of ionizing photons 3~10$^{51}$~ph
s$^{-1}$, effective temperature 45\,000~K, density 100 \cmcub and
filling factor  0.03).
}
\end{figure}

Fig. 3 shows $C_l$ as a function of the usual metallicity indicator
   \Ostrl. The meaning of the symbols is the same as in Fig. 2.
When considering only the filled symbols, there  is a weak
correlation in the sense
that \Ostrl\ tends to be smaller for larger values of $C_l$.
The correlation virtually disappears when including the open symbols,
for which  the extinction and
   reddening corrected line ratios tend to be more uncertain.
These results are confirmed by Kendall's $\tau$ statistical test;
see Table 1.
We know that the relation between \Ostrl\ and O/H is double
valued, with a maximum in \Ostrl\ occuring at log O/H + 12 of about
8.2. This is illustrated in Fig. 4, which shows \Ostrl\ versus O/H as computed
   for a sequence of simple constant density models roughly
representative of giant \hii\ regions,
using the photoionization code PHOTO described in Stasi\'nska \&
Leitherer (1996).
Because radial abundance
gradients are known to exist in spiral galaxies from the study of
their bright giant \hii\ regions (Vilas-Costas \& Edmunds, 1992,
Zaritsky, Kennicutt \& Huchra 1994), it is conceivable that, depending on the
abundance gradients and the relative weights of the regions of various
abundances in the integrated emission line fluxes, \Ostrl\ might not
be a a good indicator of the galaxy overall metallicity.
It is important to note that metallicity here refers only to the
star formation regions. Kobulnicky,
Kennicutt \& Pizagno (1999) have addressed this problem by simulating
the global \Ostrl\ of 22 spiral galaxies, integrating the observed
values for several bright individual \hii\ regions over the \Ha\
radial profiles from CCD images. They found that the oxygen abundance
derived from the simulated integrated \Ostrl\ was identical, within
errors, to the
characteristic abundance of galaxies defined by Zaritsky,
Kennicutt \& Huchra (1994) as the oxygen abundance computed at a
certain characteristic galactocentric radius. They concluded that the beam
smearing effect from sampling large number
of galaxies, even in the
presence of strong abundance gradients, has a small effect on
characterizing the mean abundances of galaxies. They mention however,
that it would be useful to test this conclusion directly using actual
integrated spectra of galaxies. Unfortunately, none of the galaxies in
their sample belongs to the NFGS sample. Our conclusion, using the
NFGS sample is purely empirical, and
is that \Ostrl\ does not seem to be a
useful indicator of metallicity in integrated spectra.

\begin{figure}
\centerline{\psfig{figure=ms1187f5.eps,width=8cm}}
\caption{
Same as Fig. 4, but for \Nii/\Ha.
}
\end{figure}

\begin{figure}
\centerline{\psfig{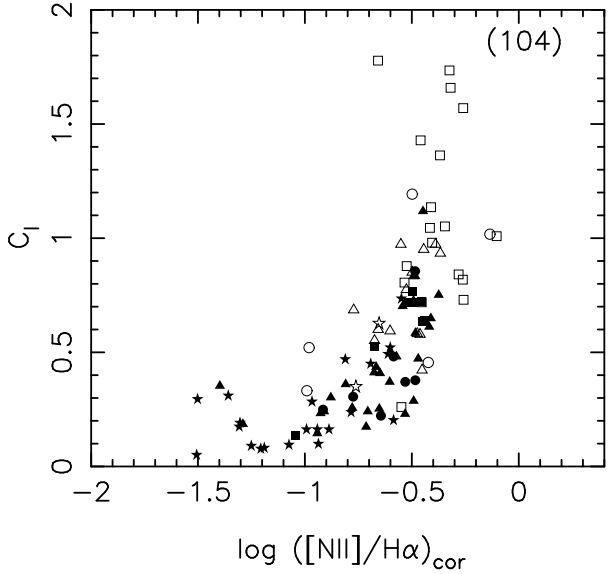}}
\caption{$C_l$ as a function of
\Nii/\Ha\
corrected for reddening. Same symbols as in Fig. 2.
}
\end{figure}

\begin{figure}
\centerline{\psfig{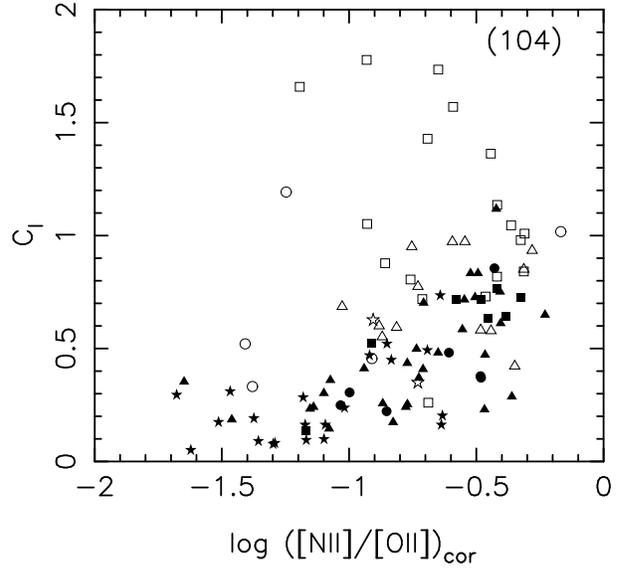}}
\caption{$C_l$ as a function of
\Nii/\Oii\
corrected for reddening. Same symbols as in Fig. 2.
}
\end{figure}

Another indirect empirical metallicity indicator has been proposed
more recently (van Zee et al. 1998a), which has the double
advantage of not being  plagued by the double value problem
and of being independent of reddening. This is the \Nii/\Ha\
ratio. Fig. 5, obtained from the same models as Fig. 4 (these
models assume that  N/O $\alpha$ (O/H)$^{0.5}$)
shows that \Nii/\Ha\
increases with O/H up log O/H + 12 around 8.8. There are two
reasons for such a behaviour. One is that \Nii/\Ha\ is less
dependent on the electron temperature than \Ostrl, so that it is not
so much affected by the decrease in electron temperature provoked by an
increase in metallicity.  The other reason is that N/H increases with
O/H. The study of giant H II regions in
spiral galaxies (van Zee, Salzer \& Haynes 1998b) indicates that at 
log O/H + 12 $>$ 8.5,
N/O is proportional to O/H and not to (O/H)$^{0.5}$,
so that the effect is even stronger than in the models shown in Fig. 5.

Fig. 6 shows $C_l$ as a function of \Nii/\Ha\ for the normal galaxies
in the NFGS sample.  There is a very clear correlation!
Kendall's $\tau$ is 0.6 and the small value of $prob$ indicates that
the correlation is highly significant.
As a matter of fact, it is even surprising that such a good
correlation should exist. Indeed, the \Nii/\Ha\ ratio obviously
strongly depends on the ionization parameter of the H II
regions. While giant H II regions seem to follow a very narrow
sequence in ionization parameters (McCall, Rybski \& Shields 1985; 
Dopita \& Evans
1986; Stasi\'nska, Schaerer \& Leitherer 2001), one expects in integrated
spectra of spiral galaxies a non negligible
contribution from a diffuse medium, possibly ionized by photons
leaking out from bona fide giant H II regions.  Indeed,
about 20 to 50\% of the H-Balmer line emission
in nearby spiral galaxies arises from this diffuse medium (e.g.,
Greenawalt, Walterbos \& Braun 1997; Collins et al. 2000; Rossa \&
Dettmar 2000; Zurita, Rozas \& Beckman 2000; Rand 2000). The \Nii/\Ha\
ratios observed in this diffuse medium are typically a factor 2 larger
than in the nearby giant \hii\ regions, which implies that in an
integrated spectrum, the effect of the diffuse medium at a given
galactocentric distance (and therefore metallicity) is to increase the
\Nii/\Ha\ ratio by about 30\% at most. Such an increment is
relatively modest compared to the
metallicity dependence of this ratio and explains why \Nii/\Ha\
appears to be such a good indicator of the overall metallicity of
spiral galaxies.  We
should emphasize that, nevertheless, this indicator
   cannot be easily translated into a value of O/H,
in the presence of abundance gradients and integration over the
complex interstellar medium of spiral galaxies.
For example, the weight of the regions of metallicities around solar
is expected to be important, as seen from Fig. 5.
One should certainly
refrain from using the calibration that has been proposed by van Zee
at al. (1998a),  which is valid only in the context of giant \hii\
regions.

Another
potential empirical indicator of metallicity of the emission line regions
which is independent of
the ionization parameter, does not suffer from the double
value problem and is less affected by  metallicity gradients is
\Nii/\Oii\ (Dopita et al. 2000). It is
however strongly dependent on the reddening correction.  Fig. 7 shows
$C_l$ as a function of \Nii/\Oii\
(the open symbols have the same meaning as in Figs. 2, 3 and 6),
and here again we clearly see a
correlation. It is more dispersed than the
$C_l$ versus \Nii/\Ha\ relation, possibly  because of errors in the
reddening correction. Indeed, the correlation is much better
   when considering only the filled symbols. This is confirmed by the
results of Kendall's $\tau$ test presented in Table 1.

  From the above considerations, our conclusion is that, indeed, $C_l$ 
correlates
with the global metallicity of the emission line regions in spiral galaxies,
at variance with the conclusions of
Zaritsky, Kennicutt \& Huchra (1994).
It should be stressed
that $C_l$ does not necessarily measure the true extinction of the total light
emitted in the Balmer lines. In the case of a patchy dust
distribution, it measures, so to say, the effective extinction of
those regions that are not too strongly obscured, so that the interpretation of
our result is by no means straightforward.

\section{The emission line sequence}

In SS99, we showed that, when ordering the normal galaxies from the
Kennicutt sample by galaxy spectral type, emission line equivalent
widths and emission line ratios from the integrated galaxy spectra
formed a very well defined sequence (not necessarily monotonic!). We
interpreted this as an evidence for a close link between the high mass
end of the stellar populations, which drive the emission lines, and
the population of the stars that are responsible for the optical
continuum, which determines the galaxy spectral type. The scatter in
the properties of individual \hii\ regions observed in spiral galaxies
(Kennicutt \& Garnett 1996, Roy \& Walsh 1997) is likely to be
   smoothed
out in integrated galaxy spectra if the galaxies are sufficiently
massive to contain a large number of \hii\ regions.
The sequence found by SS99 was so impressive, that it calls for
further investigation and the
   NFGS sample is adequately large for  such a  purpose.
   Unfortunately, we cannot determine the galaxy spectral type
from the published data, since the spectra are not yet publicly
available in digital form.  As a proxy of the galaxy spectral type, we
can use galaxy colours, with the advantage  that they are
much easier to obtain than spectral types. We found empirically that
the best colour to use was the synthetic $(B-R)$ derived from the
galaxy spectra as explained by Jansen et al. (2000b).

\subsection{Equivalent widths and star formation}

\begin{figure*}
\centerline{\psfig{figure=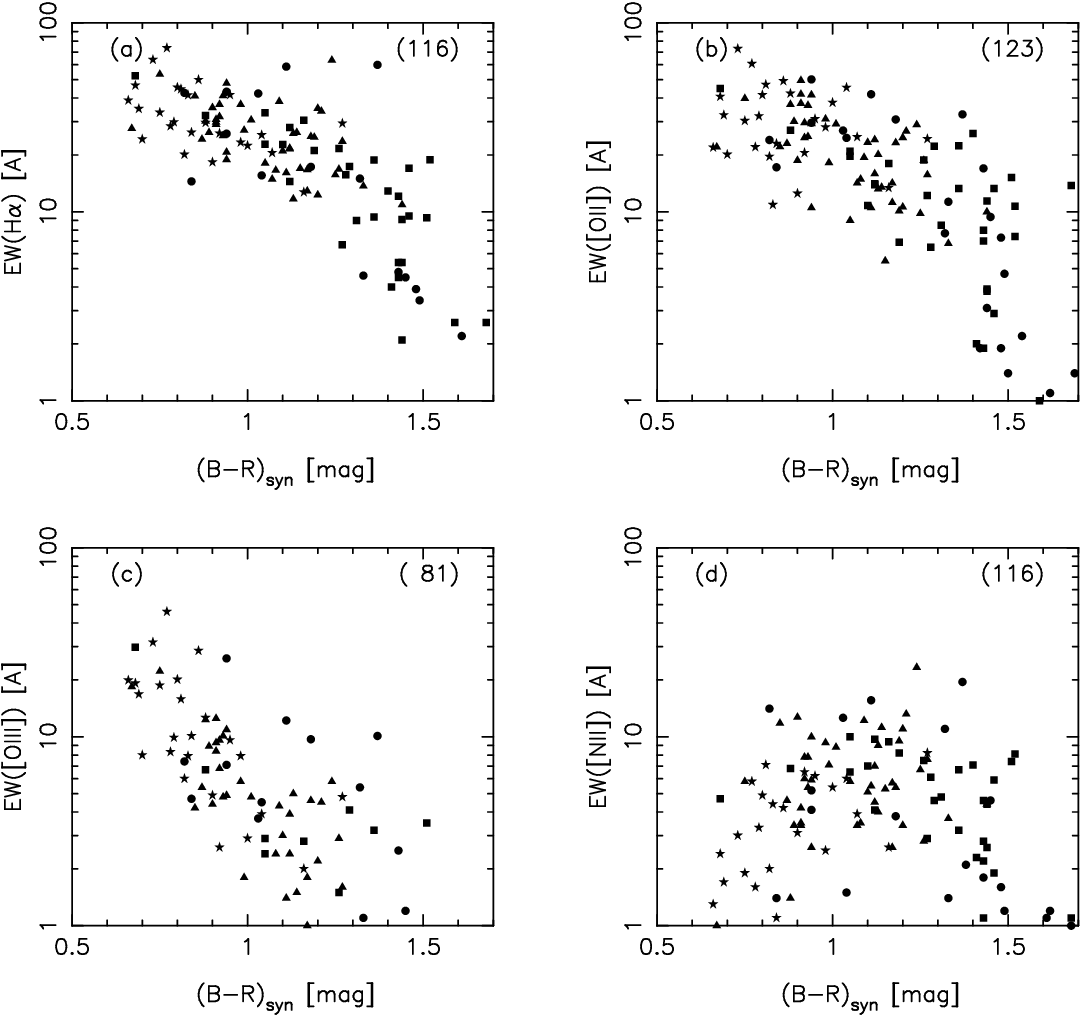,width=18cm}}
\caption{Emission line equivalent widths as a function of
the galaxy colour $(B-R)$. a) \Ha; b) \Oii; c) \Oiii; d) \Nii.
Same symbols as in Fig. 1.
}
\end{figure*}

In Fig. 8, we show the equivalent widths of \Ha, \Oii, \Oiii\ and
\Nii\ as a function of $(B-R)$.
The shape of the symbols represent the galaxy morphological types,
as in the previous observational diagrams.
Clearly, the equivalent  widths of \Ha, \Oii\ and \Oiii\ are
strongly correlated with $(B-R)$, being larger for bluer galaxies.
However, there seems to be more dispersion than
what was suggested by the study of SS99. One possible reason is that
the color $(B-R)$ is not as good a parameter as the spectral type for
ranking galaxies.
For comparison, we show in Fig. 9 the
same equivalent widths but as a function of the galaxy morphological
type, $T$. The correlations of equivalent widths with $T$ are seen to be
much weaker than with  $(B-R)$.
These conclusions are confirmed by the statistical test summarized
in Table 1.

\begin{figure*}
\centerline{\psfig{figure=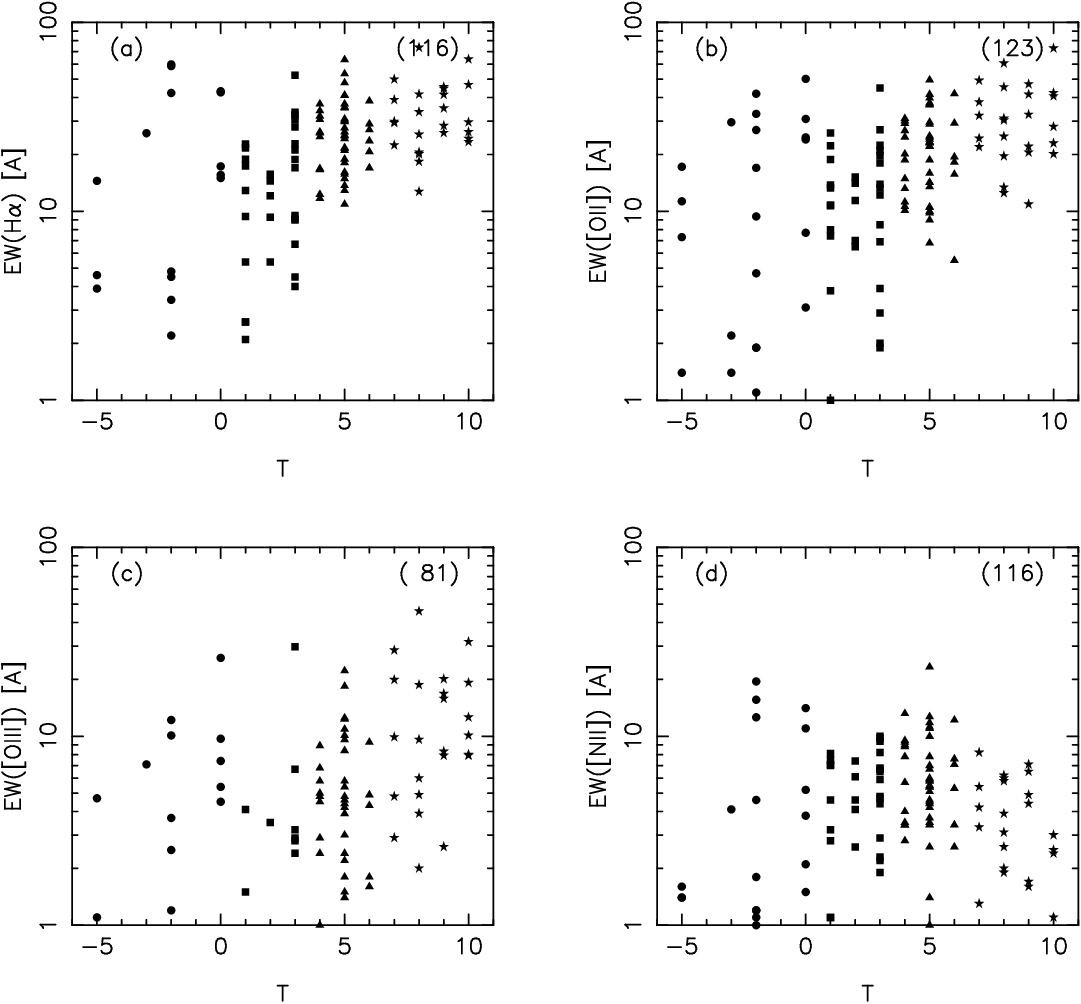,width=18cm}}
\caption{ Emission line equivalent widths as a function of
the galaxy morphological type $T$. a) \Ha; b) \Oii; c) \Oiii; d) \Nii.
Same symbols as in Fig. 1.
}
\end{figure*}

Emission lines, principally \Ha\ and \Oii,  are indicators of current star
formation (Barbaro \& Poggianti 1997, Kennicutt 1998,
Jansen, Franx \& Fabricant 2001).  The calibration in terms of star
formation rates however strongly depends on several factors:
star formation law, metallicity, selective dust extinction,
etc. Charlot \&
Longhetti (2001) argue for the necessity of using the
\Oii, \Oiii\, \Ha\, and \Hb\ lines at the same time to better constrain the
problem. Qualitatively, though, Fig. 8 (together with Fig. 1) shows
that any of these
parameters is likely to be a first order ranking indicator of
integrated star formation
rates in spiral galaxies. This is of course not the case of \Nii,
whose equivalent width behaves in
a completely different manner. Indeed, the strong drop in nitrogen
abundance for the bluest (less chemically evolved) galaxies overtakes
the effects
of increasing present star formation rate and increasing electron
temperature.

As shown by Jansen et al. (2000b), equivalent widths are also loosely
correlated with galaxy absolute magnitudes, tending to be larger for more
luminous galaxies. However, at a given galaxy colour, we found no
compelling evidence of brighter galaxies having larger equivalent
widths.

\subsection{Line ratios and metallicity}

\begin{figure*}
\centerline{\psfig{figure=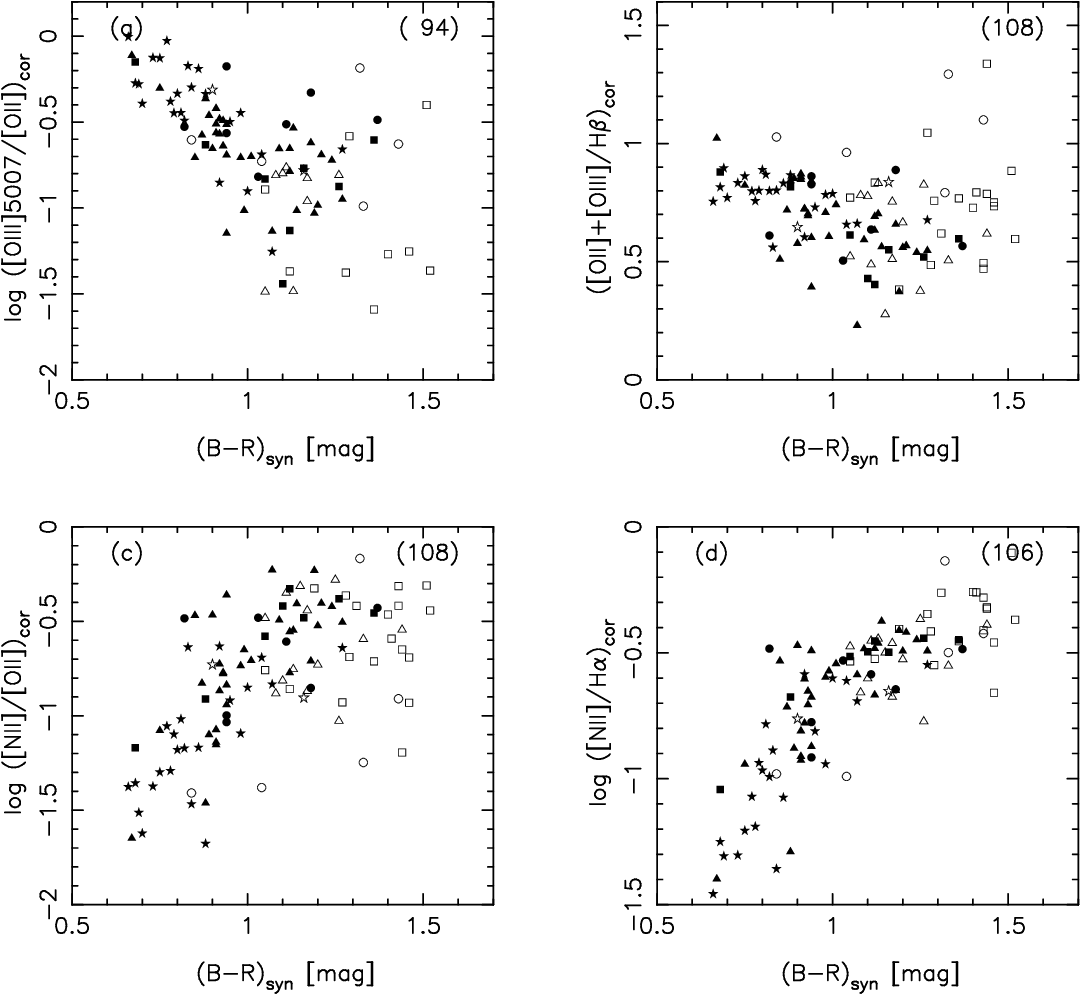,width=18cm}}
\caption{Emission line ratios corrected for reddening  as a function of
the galaxy colour $(B-R)$. a) \Oiii/\Oii; b) (\oiii + \oii)/\Hb; c)
\Nii/\Oii; d) \Nii/\Ha.
Same symbols as in Fig. 2.
}
\end{figure*}

Fig. 10 shows various emission line ratios as a function of
$(B-R)$. The line ratios have been dereddened using the
value of C$_{l}$ computed above. Since the reddening correction becomes
more uncertain for galaxies with small \WHa\ and \WHb, as in Figs. 2, 3, 6 \& 7
we marked  by filled
symbols those objects that have \WHa $>$ 10~\AA\ and  \WHb $>$ 5~\AA\ and thus
the most reliable reddening
corrections.  Clear trends are seen for \Oiii/\Oii,
\Nii/\Oii\ and \Nii/\Ha\ as a function of $(B-R)$. As for the
equivalent widths,
we find the trends much clearer with $(B-R)$ than with morphological type.
For \Oiii/\Oii\ and \Nii/\Oii, which are sensitive to
reddening, one can see that the open symbols prolongate the trends
defined by the
filled symbols, but with much higher dispersion. In the case of
\Ostrl, there is a weak trend with $(B-R)$ when considering only the
filled symbols, and practically no trend when considering the entire data set.
The statistical significance of these trends are presented in Table 1.

The fact that \Oiii/\Oii\ decreases as the galaxy color gets redder is not
necessarily only due to a lowering of the average ionization parameter
or the average spectral hardness of the exciting stars (as was
suggested by SS99). Indeed, an increase in overall metallicity, by
enhancing the temperature drop in the central zones of \hii\ regions
with metallicities larger than solar, also contributes to lower the
\Oiii/\Oii\ ratio (see Stasi\'nska, Schaerer \& Leitherer 2001). We 
find no trend of
\sii/\Ha\  with colour, suggesting
   that the effect of ionization parameter
along the colour sequence may not be overwhelming.

As mentioned in Sect. 2, the \Ostrl\ ratio is not the best indicator
of metallicity in integrated spectra of galaxies where abundance
gradients are known to occur. Being non monotonic with respect to the
metallicities of individual \hii\ regions, when integrated over an
entire galaxy it becomes impossible to interpret without additional
information on the radial galactic gradients in abundances and star formation
rates. In addition, it is strongly dependent on the reddening correction,
which becomes uncertain in metal rich galaxies with weak lines.

On the other hand,  \Nii/\Oii\ is a good indirect indicator of
metallicity in extragalactic \hii\ regions (Dopita et al. 2000) and is
expected to be a fair indicator of overall metallicity in integrated
galaxy spectra (see above).  The trend seen in Fig. 10c is
produced by  an
increase of N/O as galaxy colour gets redder, indirectly
indicating an increase in metallicity. As a consequence, the
conspicuous trend of \Nii/\Ha\ increasing with $(B-R)$ (Fig. 10d),
is likely due to an increase in metallicity. As mentioned before, the
\Nii/\Ha\ ratio has the advantage of being reddening independent, and
it can be used with confidence even for galaxies with weak emission
lines. It clearly indicates that the sequence between metallicity and
$(B-R)$ colour is quite narrow even for galaxies redder than $(B-R)$=1.2
(which is not apparent when using the strongly reddening-dependent
\Nii/\Oii\ ratio).

\begin{figure*}
\centerline{\psfig{figure=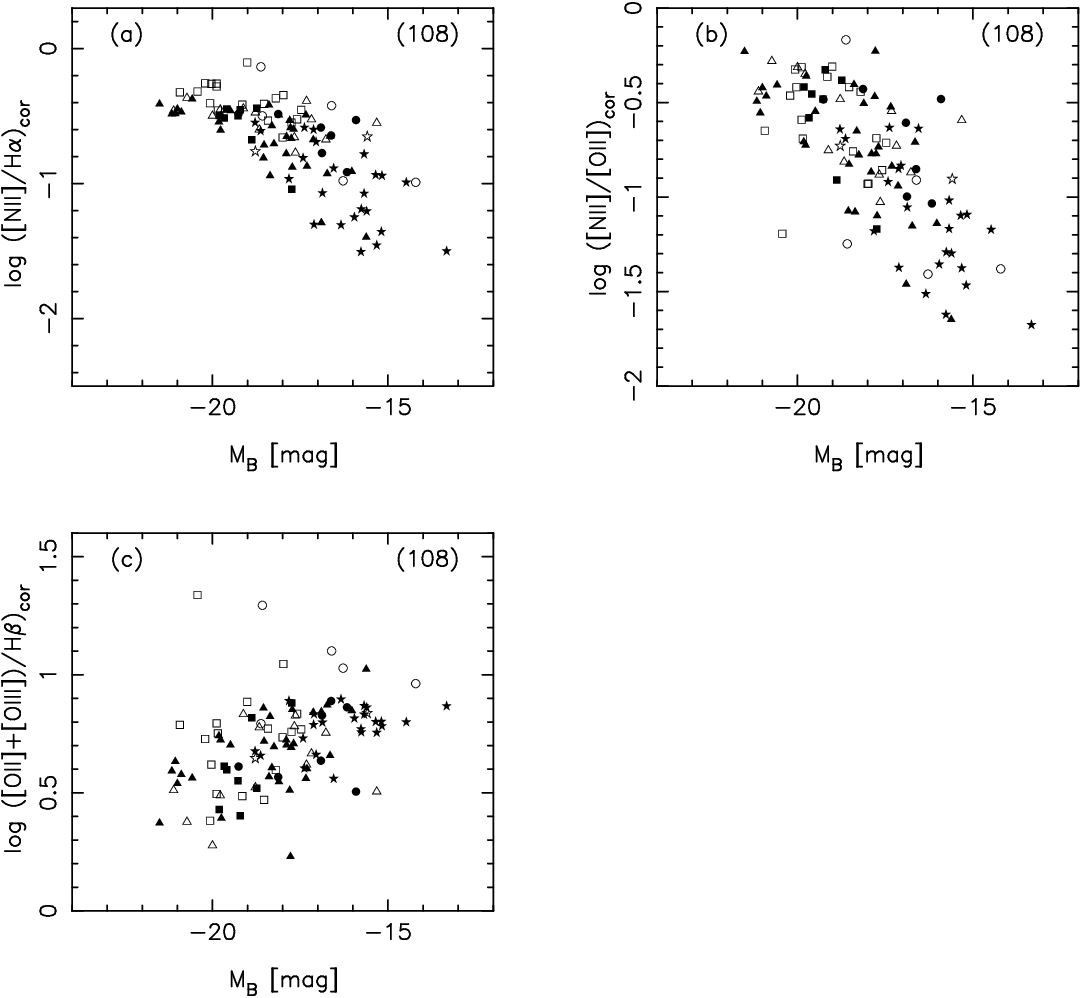,width=18cm}}
\caption{Emission line ratios corrected for reddening  as a function of
the galaxy absolute blue magnitude$M_B$.  a) \Nii/\Ha; b) \Nii/\Oii;
c) (\oiii + \oii)/\Hb.
Same symbols as in Fig. 2.
}
\end{figure*}

Numerous studies have claimed the existence of a
metallicity-luminosity relation in a variety of classes of galaxies:
dynamically hot galaxies, i.e. ellipticals, bulges, and dwarf
spheroidals, dwarf \hii\ galaxies, irregular galaxies and spirals
(Bender et al. 1993; Vigroux, Stasi\'nska \& Comte 1987; Skillman,
Kennicutt \& Hodge 1989; Richer
\& McCall 1995; Zaritsky, Kennicutt \& Huchra 1994). In Fig. 11, we show the
various metallicity indexes, \Ostrl, \Nii/\Ha\ and \Nii/\Oii\ as a
function of the total absolute blue magnitude $M_B$. We do find a
strong correlation between $M_B$ and the metallicity indexes
\Nii/\Ha\ and \Nii/\Oii.
The correlation of the \Ostrl\ index with $M_B$ is
statistically significant (see Table 1), but at a rather low level.
It is interesting to recall
that Zaritsky, Kennicutt \& Huchra (1994), by using this index to infer the
characteristic metallicity of spiral galaxies, did find a trend with
galaxy luminosity. But the index actually referred to the observed
ratio at a certain galactic radius, and not to the integrated light
from the galaxy. Our study has shown how misleading can be
the use of \Ostrl\ in integrated spectra of galaxies in the presence
of abundance gradients. On the other hand, \Nii/\Ha\ and \Nii/\Oii\
show clear trends with galaxy absolute magnitude, confirming that
indeed, there is a relation in spiral galaxies between the overall
metallicity of the
star forming regions and the galaxy luminosity.

\section{Principal Component Analysis}

Principal Component Analysis (PCA) is a useful tool for the examination
of multidimensional data and the identification of underlying variables
that may be responsible for the variance in a data set (Murtagh \& Heck
1987, Fukunaga
1990). PCA allows to define a new
orthonormal reference system in a parameter space, with basis-vectors
(the principal components) spanning directions of maximum variance.

Here we apply PCA to a set of variables describing different
aspects of the galaxies in the NFGS sample. The 7 variables selected
for this exercise are:
the blue absolute magnitude $M_B$; the color $(B-R)$; the numerical
value of the morphological type $T$; the galaxy surface brightness at
the effective radius $\mu_B^e$;
\WHa, which is related to the star formation rate;
the metallicity indicator \Nii/\Ha; and H$\alpha$/H$\beta$,
related to the emission-line extinction. The number of normal galaxies
with complete information that enters in the analysis is 102.

Since the
input variables are of different nature, it is appropriate to
apply PCA on the correlation matrix, that is, the input variables
were transformed: they had their mean value subtracted and, after, they
were divided by their standard deviation. Hence the analysis is made on a set
of transformed variables with zero mean and unity variance.
The result indicates that the first principal component explains alone
48\% of the sample variance. The second and third components are responsible
for 19\% and 12\% of the sample variance, respectively.

Although each  principal component is a linear combination of all
the input variables, it is interesting to verify whether some variables
are well correlated with the first components because, in this case,
they are responsible for a significant fraction of the sample variance.
We have then computed, for each variable and principal component, the
non-parametric Kendall's $\tau$ coefficient of rank correlation and the
respective probability of absence of correlation between these
quantities. We found that the
variable that best correlates with the first component is the
metallicity indicator \Nii/\Ha; the variables
$(B-R)$, and \Ha/\Hb\ also correlate well
with the first component; the  morphological type is  correlated with the
first principal component, but to a lesser extent. On the other side,
only $\mu_B^e$ is
strongly  correlated with the second principal component. The absolute
magnitude correlates best with the third principal component.
Hence, this analysis indicates that the variance in the parameter space
considered here is produced, in first place, by the metallicity and
parameters related to it or to the stellar populations, and, in second place,
by the surface brightness, which is related to the dynamical
history of the galaxies. The absolute magnitude, related to
the mass of the galaxy, surprisingly comes only in the third place.

\section{Conclusions}

We have explored the  emission line trends from the integrated spectra
of normal spiral galaxies, in order to investigate the relationships between
dust extinction, metallicity and some macroscopic properties of spiral
   galaxies. A similar aim was followed by the study by Zaritsky, 
Kennicutt \& Huchra (1994)
   on a sample of 39 disk galaxies, which however was not based on
integrated spectra of galaxies, but on the study of several
giant \hii\ regions in each galaxy.
Our approach, using integrated spectra, is a priori likely
   to give more robust answers. It also has the advantage of
   being readily applicable to
studies of galaxies at high redshift.
We benefited from data from the Nearby Field Galaxy Survey
(Jansen et al. 2000a and b), which considerably increased the
size of the so far available samples for systematic studies
   of the global properties of spiral galaxies.

Our main results are the following.

We found that the correlation between  \WHa\ and \WHb\ is extremely strong.
This implies that $C_l$ -$ C_c$, the difference between the effective
extinction of the stellar and the nebular light,
   depends only on the intrinsic colours of the galaxies,
being larger for redder galaxies.

The nebular extinction as derived from the Balmer decrement
   is found to correlate essentially with the effective metallicity
of the emission line regions. This finding required
using an appropriate metallicity indicator. The usual \Ostrl\
index does not work for integrated spectra of galaxies, due to
strong metallicity gradients. On the other hand, \Nii/\Ha\ and \Nii/\Oii\
seem to be adequate  empirical metallicity indicators,
the first one having the advantage of being reddening independent and being
applicable also for galaxies with weak emission lines. The
calibration of these indicators would require a modelling of the
integrated spectra of galaxies, in the vein of the works of
Charlot \& Longhetti (2001) or Moy, Rocca-Volmerange \& Fioc (2001)
but including the effects of radial gradients in
the metallicity and star formation rates.

We find that the overall metallicity of the emission line regions,
as inferred from adequate indicators,
   is  strongly correlated with galaxy colours.
The correlation is much stronger  than with morphological types.
This suggests that in normal spiral galaxies,
the global metallicity of the star forming regions and
the old stellar populations are closely linked together,
   while morphological types are influenced by additional factors in
the galaxy histories.

In order to better evaluate the relation between the different
parameters describing
   the galaxies, we performed a PCA analysis on a 7-D parameter space.
We found that the variance in the considered parameter space
is produced, in first place, by the metallicity and parameters linked to the
stellar populations,
   and, in second place,
by the surface brightness, which is linked to the dynamical
history of the galaxies. The absolute magnitude, related to the
mass of the galaxy,  comes only in the third place.

The strong empirical evidences provided by this exploratory work bring
important constraints that will have to be reproduced by models of spiral
   galaxy formation and evolution, in order to understand what
drives the spiral galaxy sequence.

\acknowledgements

This study would not have been possible
without the huge observational work by R. A. Jansen, D. Fabricant,
M. Franx \& N. Caldwell, and we acknowledge
these authors for having made available their results on the Web.  We
benefited from a financial help from CNRS and CNPq. G.S. warmly
acknowledges the IAG and L.S. the DAEC for hospitality.
L.S. also acknowledges Fapesp and Pronex for support.


\end{document}